# WAS LEPENSKI VIR AN ANCIENT SUN OR PLEIADES OBSERVATORY?

Vladan Panković , Milan  Mrdjen, Miodrag Krmar

Department of Physics, Faculty of Sciences, 21000 Novi Sad, Trg Dositeja Obradovića 4. , Serbia, vladan.pankovic@df.uns.ac.rs

## Abstract

In this work we testify using simulation programs some old hypotheses according to which remarkable mesolithic village Lepenski Vir (6500 – 5500 BC) at the right (nearly west) Danube riverside in the Iron gate in Serbia was an ancient (one of the oldest) Sun (proto) observatory. We use method recently suggested by A. C. Sparavigna, concretely we use at internet simply and freely available software or local Sun radiation direction simulation and calculation computer programs. In this way we obtain and discuss figures of the sunrise in the Lepenski Vir during winter and summer solstice and spring and autumnal equinox in relation to position of the mountains, especially Treskavac (Trescovat in Romanian) and Kukuvija at left (nearly east) Danube riverside (in Romania). While mountain Kukuvija represents really the natural marker for the Sun in date of the winter solstice, mountain Treskavac, in despite to usual opinions, does not represent any real marker for the Sun in date of the summer solstice. Sun rises exactly behind Treskavac, relatively exactly (in the domain of used methods) speaking, between 20. April and 1. May. It more or less corresponds to year period when heliacal rising of the Pleiades constellation occurs, which by many ancient cultures, e.g.  Celts of northern Europe, denotes very beginning of the year. Really, in common with some other facts obtained using at internet simply and freely available stellarium software or sky objects position simulation programs, we demonstrate that for the Lepenski Vir observer heliacal rising of the Pleiades ends exactly at 20. April and this observationally accurately determined event can be used for very beginning of the year. All this opens a very probable possibility according to which Lepenski Vir was really an ancient (one of the oldest) Sun and Pleiades constellation (proto)observatory too with Treskavac as a (natural) marker.

## 1. Introduction

As it is well-known [1], [2], [3], [4] Lepenski Vir discovered by D. Srejović represents remarkable mesolithic village and culture (about 6500 – 5500 BC) at the right (nearly west) Danube riverside in the Iron gate in Serbia at the border with Romania. In the Lepenski Vir originated first monumental sculptures in the Europe, on the one hand, and base of the houses in the Lepenski Vir have had original, specific nearly trapezoidal forms and their buildings have been realized using very sophisticated geometrical and architectonic procedures. Nevertheless form of the basis of the house is very similar to the trapezoidal form of the mountain Treskavac (Trescovat in Romanian) placed oppositely to the Lepenski Vir at left (nearly east) Danube riverside (in Romania). It does not represent a coincidence only. On the contrary it expresses a deep connection of the culture with religious elements of the Lepenski Vir with Treskavac. It causes some less or more accurate hypotheses according to which Lepenski Vir with Treskavac have had some astronomical characteristics [1]. [2], [5] even that it was an ancient (one of the oldest) Sun (proto)observatory.

In this work we shall test such hypotheses using methods recently suggested by A. C. Sparavigna for some other ancient Sun observatories [6], [7], [8]. Concretely we shall use the following at internet simply and freely available software or different local Sun direction simulation and calculation computer programs: 1. http://www.sollumnis.com/ recomanded. by Sparavigna [6], [7], [8] , 2. http://suncalc.net/#/51.508,-0.125,2/2014.12.02/14:25 , 3. http://www.spectralcalc.com/solar_calculator/solar_position.php . Additionally, we shall use at internet simply and freely available stellarium software or sky objects position simulation program 4. http://www.fourmilab.ch/cgi-bin/Yourhorizon . All four mentioned and similar programs can be simply funded at internet by everyone which can check and affirm our results. Also, everyone can in the analogous way simply find additional data that go over basic intention of this work and that, for this reason or reason for simplicity, are here omitted.

By first and second program we shall simply obtain figures of the satellite and ground map of the Lepenski Vir and its close environment. We start programs by insertion of two necessary words, Boljetin (village about 5km southern to Lepenski Vir) , and, Serbia, after which naviation procedure leads simply toward Lepenski Vir with latitude 44.55° and longitude 22.02°.

Then we shall obtain and discuss figures of the sunrise in the Lepenski Vir during winter and summer solstice, spring and autumnal equinox , and some other astronomically potentially interesting dates and intervals, in relation to position of the mountains, especially Treskavac and Kukuvija at left (nearly east) Danube riverside (in Romania). While, according to obtained figures, top of the mountain Kukuvija represents really the natural marker for the Sun in date of the winter solstice, mountain Treskavac, in despite to usual opinions, does not represent any real marker for the Sun in date of the summer solstice.

Obtained figures clearly show that Sun rises behind Treskavac, relatively accurately (within our method) speaking, between 20. April and 1. May. It, roughly speaking, corresponds to year period when heliacal rising of the Pleiades constellation occurs, and this event by many ancient cultures, e.g. Celts of northern Europe, denotes very beginning of the year [9], [10].

For reason of a relatively accurate analysis of the correlations between time interval of the sunrise immediately behind Treskavac and Pleiades constellation heliacal rising we shall use mentioned third and fourth programs Namely, according to the third program sunrise moment at Lepenski Vir at any day in the year, including days between 20. April and 1. May, can be exactly determined by calculations (al this can be doe by first and second program too). This moment will be introduced in the fourth program which will simulate figure of the sky horizon (with Pleiades constellation position) in this moment for days between 20. April and 1. May.

All this will open a very probable possibility that Lepenski Vir was an ancient Sun and Pleiades constellation (proto)observatory with Treskavac as a (natural) marker.

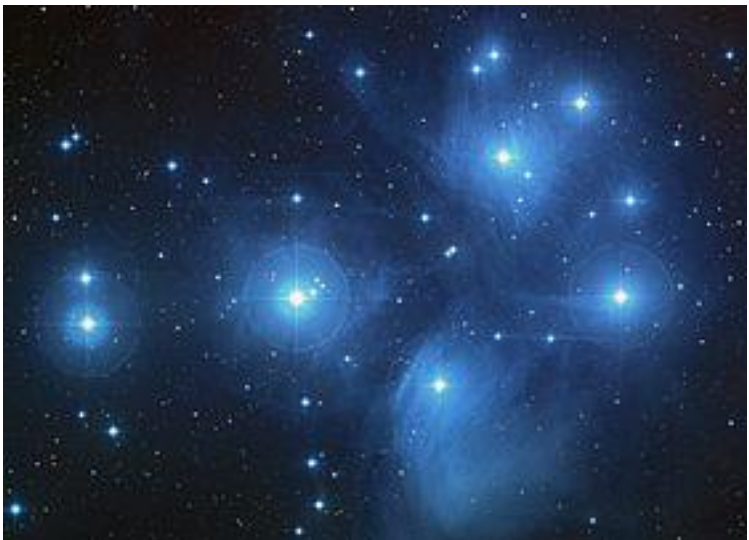
Fig. 1  Pleiades constellation

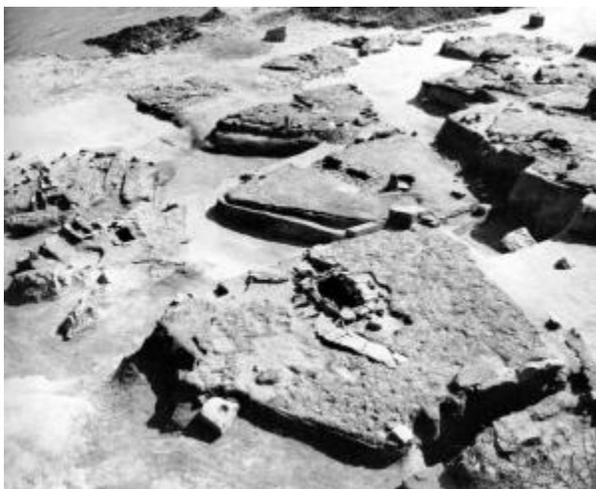
Fig. 2  Bases of the houses in Lepenski Vir  – photography from [1], [2]

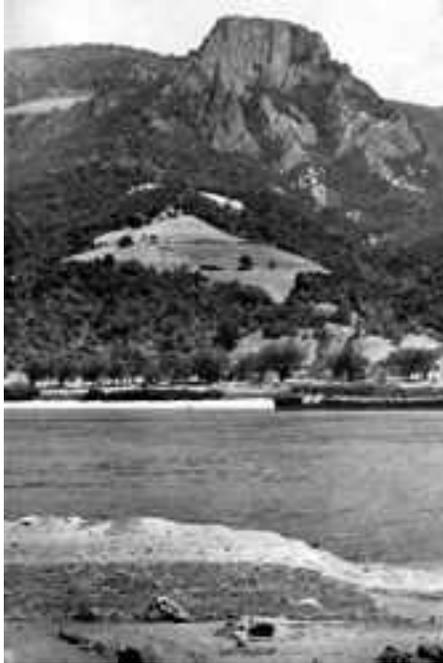

**Fig. 3  Basis of the house in Lepenski Vir opposite to Treskavac – photography from [1], [2]**

## 2. Lepenski Vir an ancient Sun and Pleiades observatory – consequence of the simulated figures

We introduce two mentioned words, Boljetin , and, Serbia, in mentioned first or second program, 1. http://www.sollumnis.com/  and 2. http://suncalc.net/#/51.508,-0.125,2/2014.12.02/14:25. After navigation, we obtain figure 4 of the satellite map of Lepenski Vir with close environment. More detailed analysis by the same programs, which here will not be given explicitly, points out that Lepenski Vir holds latitude 44.55° and longitude 22.02°. At this figure 4 Lepenski Vir overlapped village area ,with represents small white spot at the left hand of the figure or right Danube riverside opposite to white bare rock, i.e. mountain Treskavac. (The following here can be pointed out. After building of the hydropower Djerdap I, whole original plateau of the Lepenski Vir village has vertically lifted and covered for reason of the protection from atmospheric impacts. Nevertheles this lifting and overlap has no significant influence on the observations referring at Lepenski Vir ancient observatory hypotheses, since linear dimensions of the Lepenski Vir area are much smaller than distance between the Lepenski Vir and mentioned nearly mountains.)

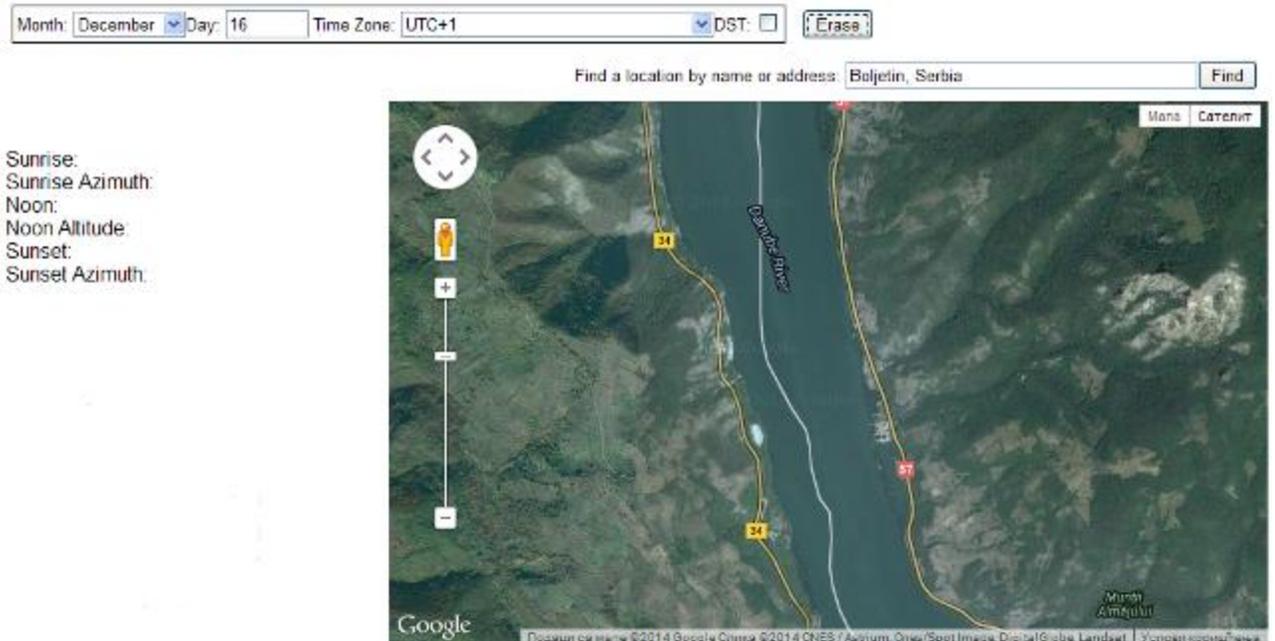

**Fig. 4**

Then we decrease domain of the observation and focus our attention at the overlapped area of the Lepenski Vir village only. We chose a point on this overlapped white area and chose the date of the winter solstice (22. December) after which sunrise and other characteristic directions become created using programs (with an implicit notation of the latitude and longitude). All this is presented at figure 5.

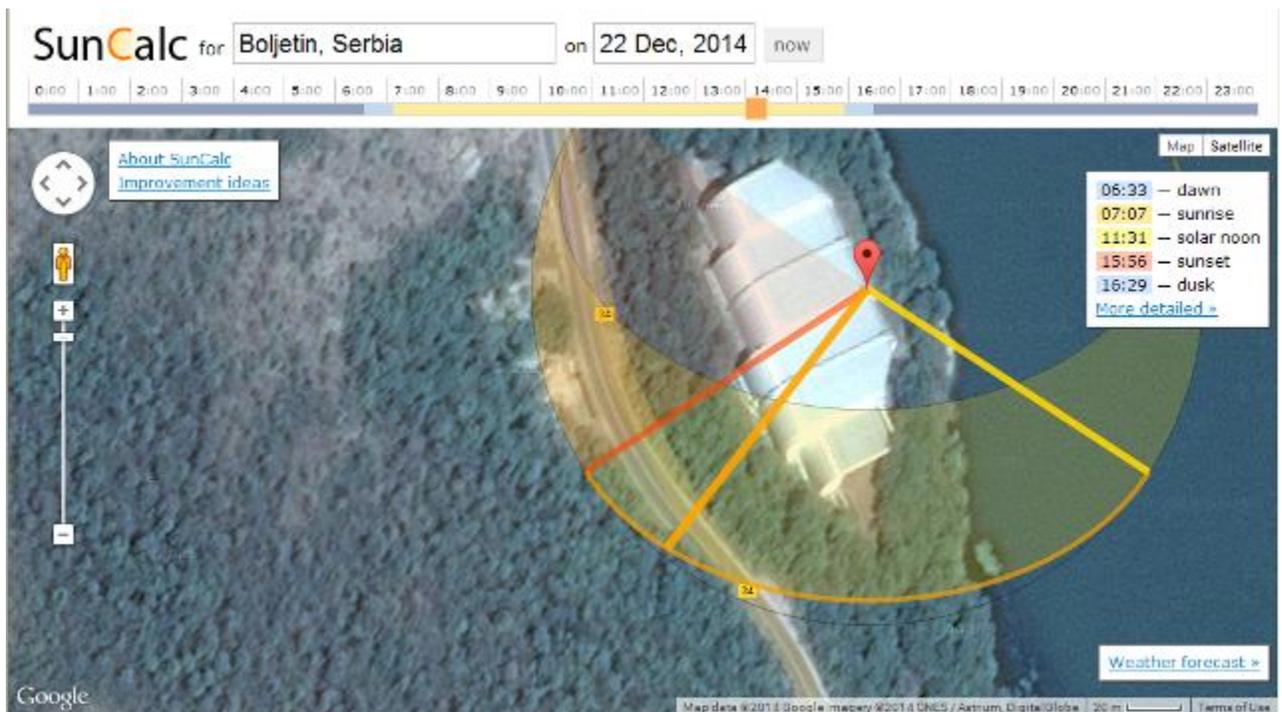

**Fig. 5**

Further we change previous satellite map by ground (street) map at which heights of the locations are relatively clearly denoted using isohypses (contours) method. Now overlapped Lepenski Vir village becomes immediately effectively unobservable but it becomes intermediately observable using endpoint of the sunrise line. Also we increase the domain of the observations and recognize unique hill Kukuvija (immediately beside Danube left riverside) behind of which Sun rises for Lepenski Vir observer at winter solstice (22. December). All this is presented at figures 6 and 7.

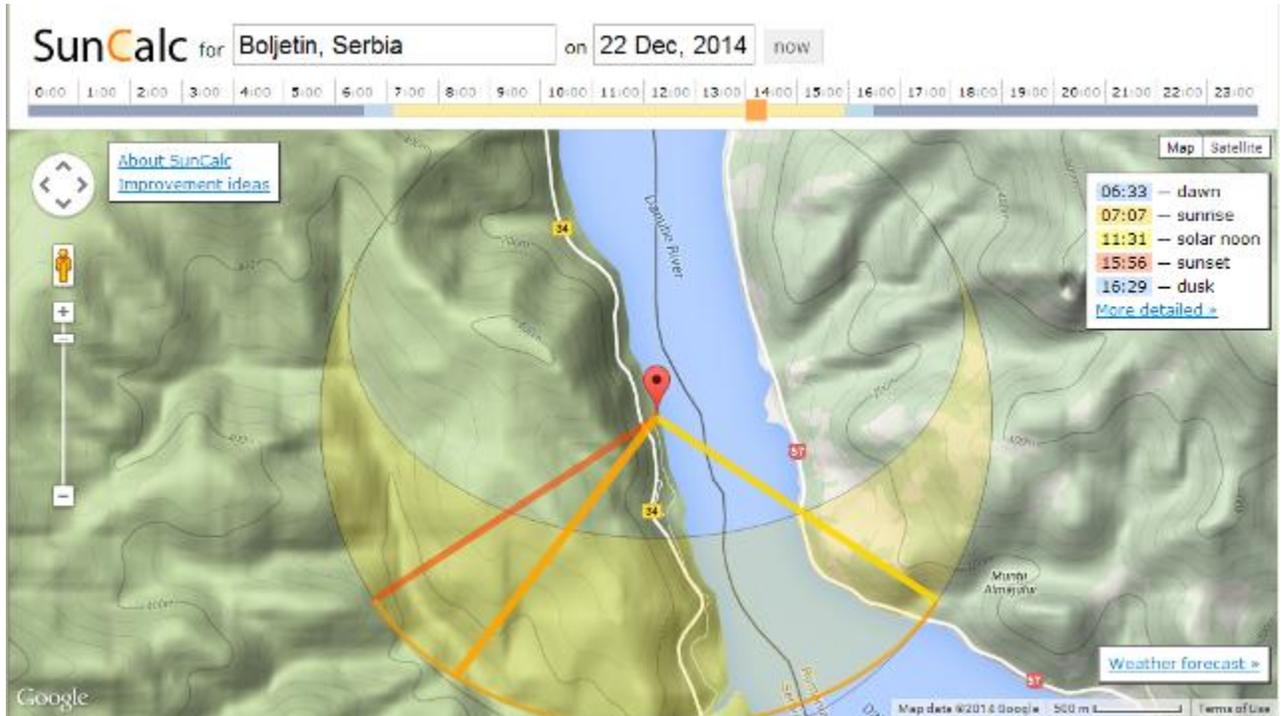

**Fig. 6**

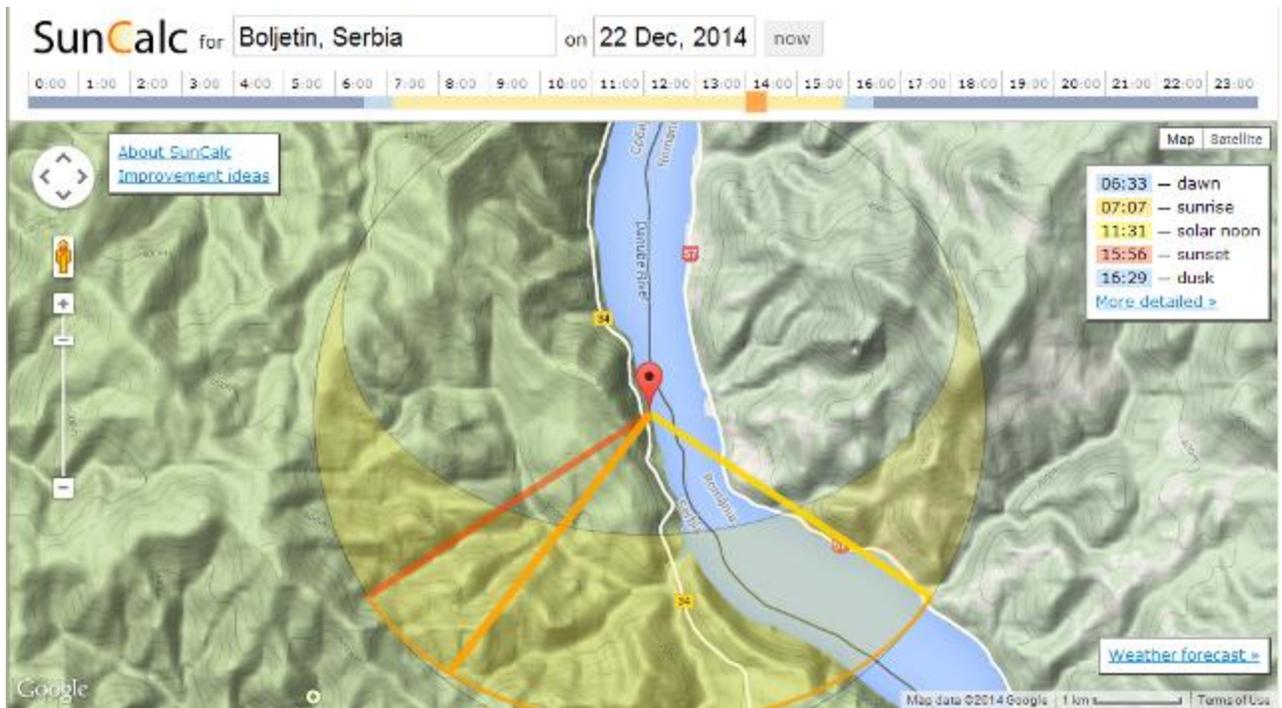

**Fig. 7**

By date change we can, using mentioned programs, do figures 8 and 9 at which we can recognize unique mountain or hill behind which Sun rises for Lepenski Vir observer at spring (21. March) and autumnal (23. September) equinox.

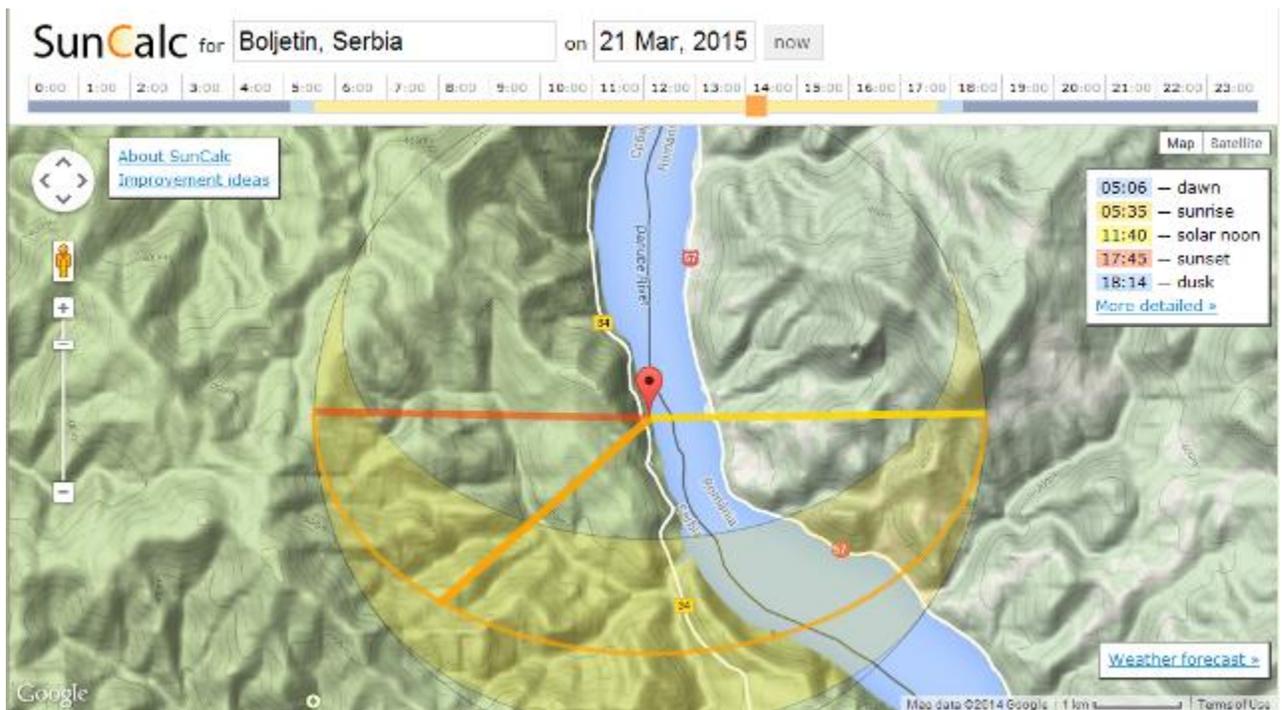

**Fig. 8**

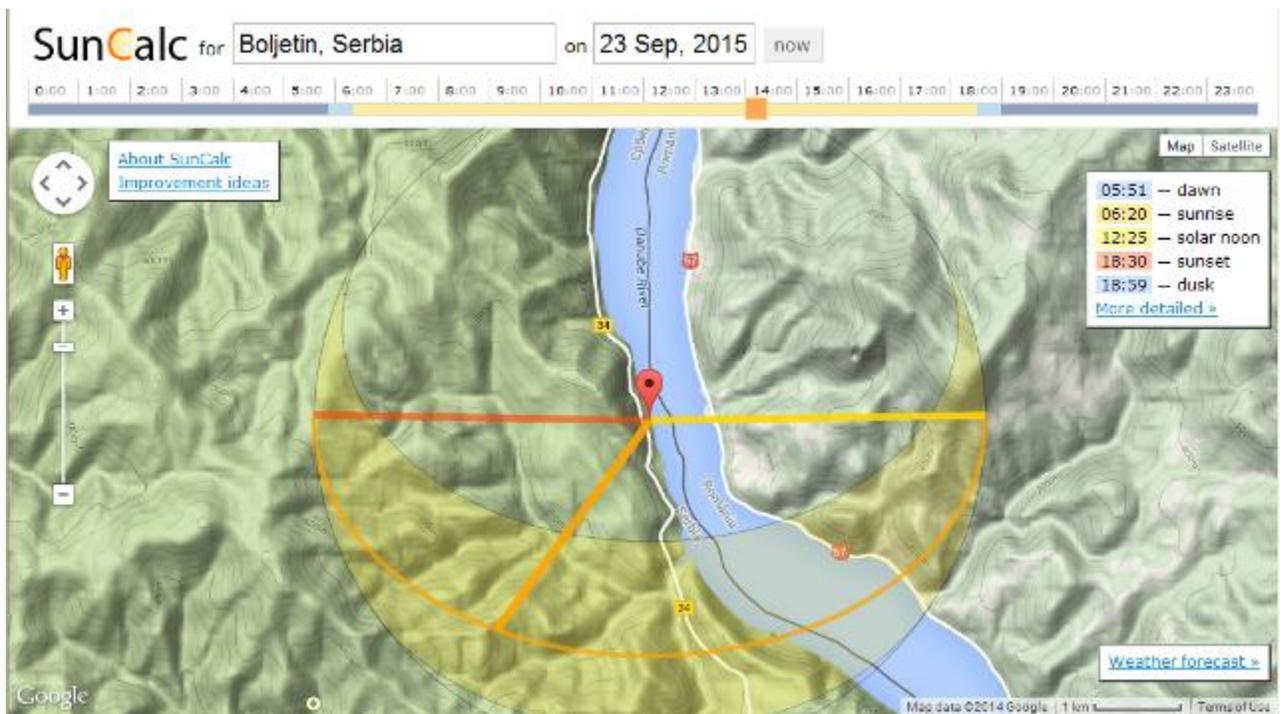

**Fig. 9**

Further, by date change, we can, using mentioned programs, do figures 10. and 11. which refer on the sunrise for Lepenski Vir observer at summer solstice (22. June). Quite obviously neither Treskavac nor any other mountain or hill top represents the place behind which sunrise occurs for Lepenski Vir observer at summer solstice. For the Lepenski Vir observer sunrise at summer solstice occurs behind a mountain ridge point roughly speaking northern in respect to Treskavac and with the same height as the Treskavac base.

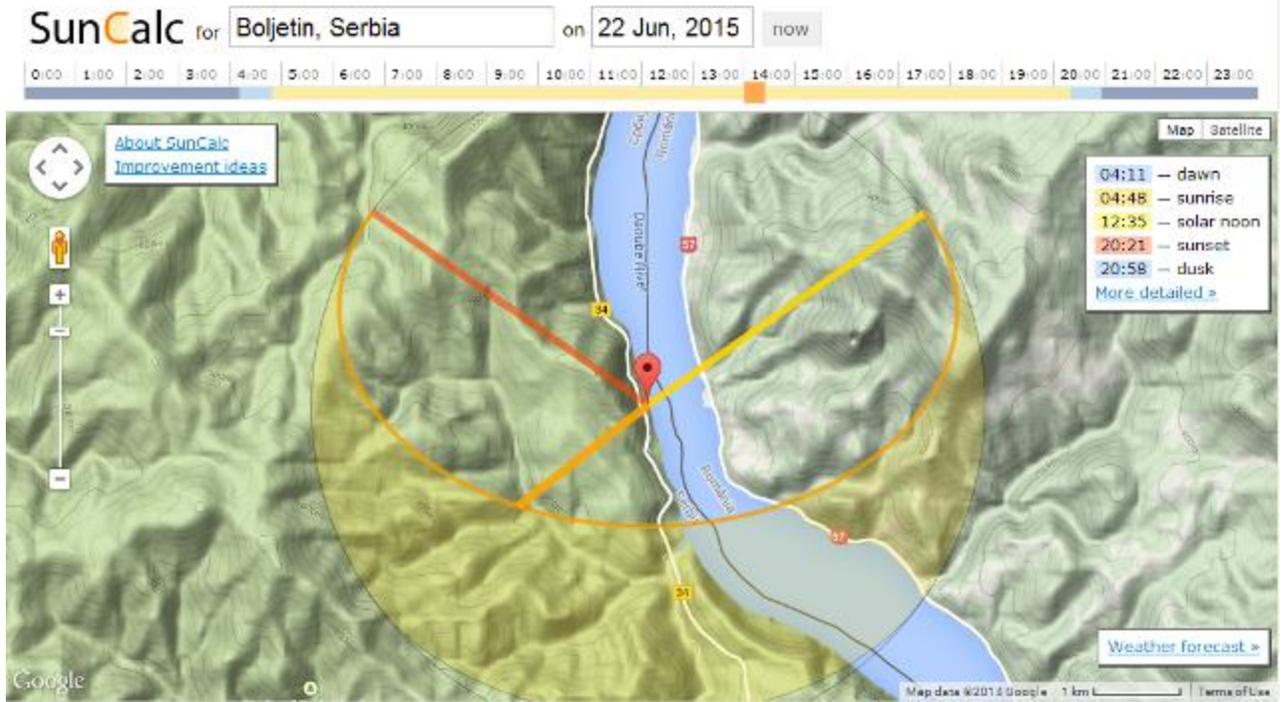

**Fig. 10**

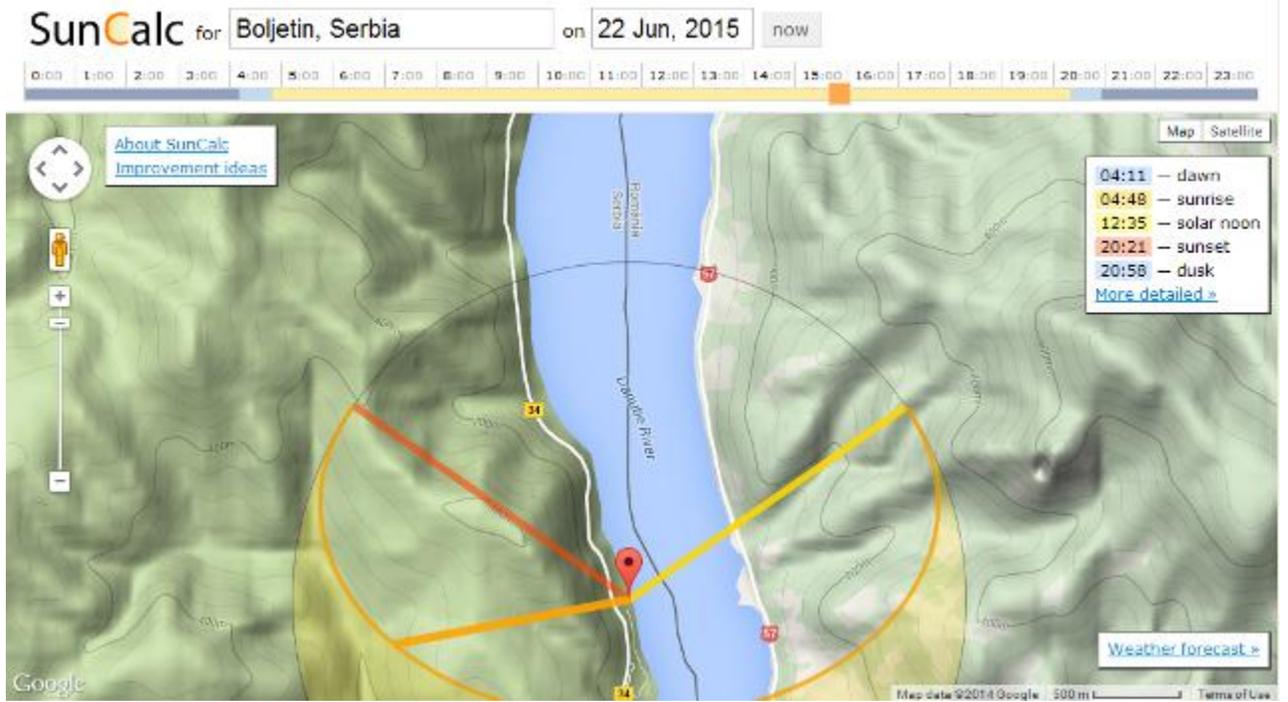

**Fig. 11**

Now we can, according to all previous figures 6. – 11. , practically uniquely determine such locations nearly Lepenski Vir that can be used for the markers for the sunrise in winter solstice, spring equinox, summer solstice and autumnal equinox. These locations are marked at figure 12. (implicitly) and 13 (explicitly).

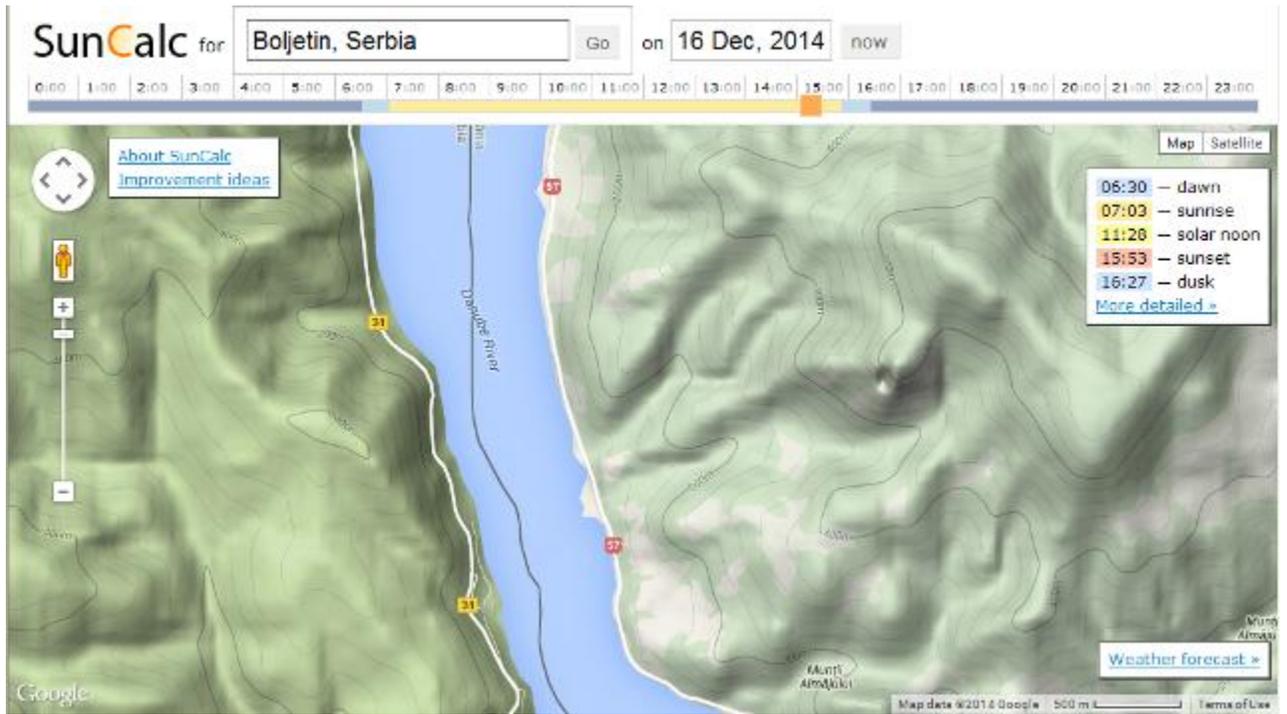

**Fig. 12**

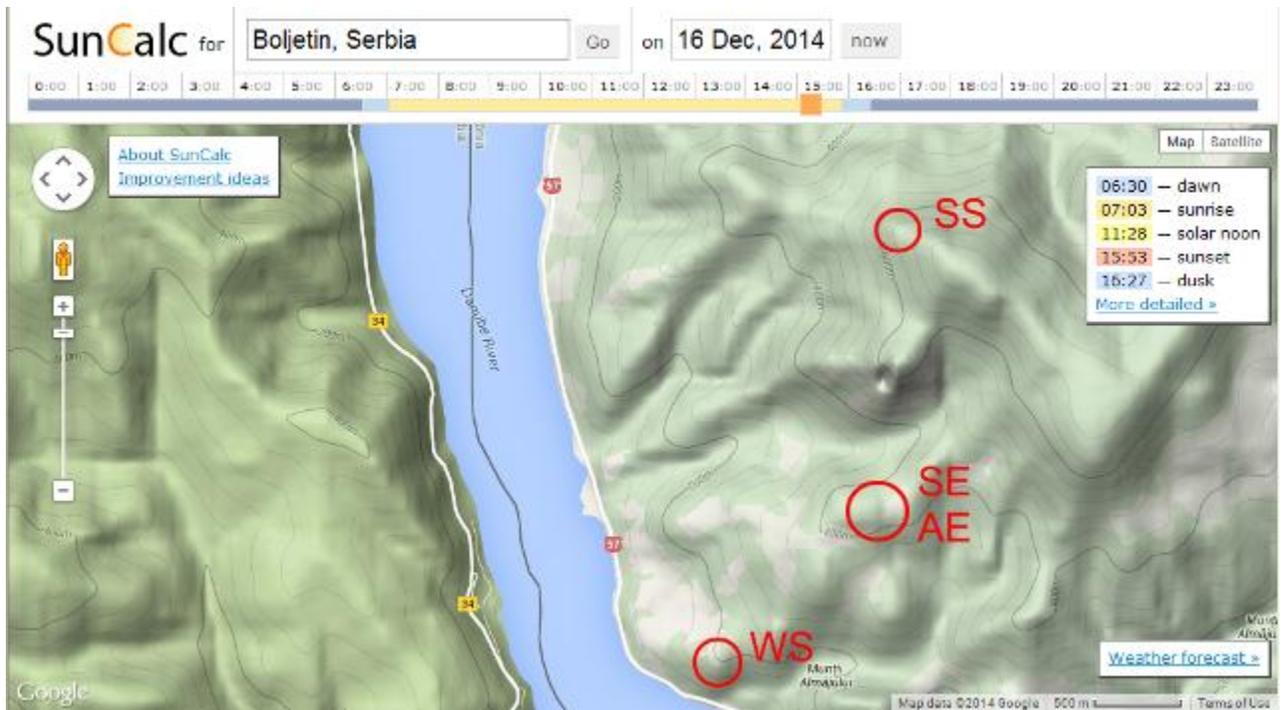

**Fig. 13 – SS (summer solstice sunrise point), SE and AE (spring and autumnal equinox sunrise point), WS (winter solstice sunrise point)**

Even if Treskavac does not represent any of previously noted markers, it, according to unambiguous archeological facts [1], [2] represents probably most significant mountain for the Lepenski Vir people. For this reason we shall analyze using mentioned programs, figures and date changes, time period when sunrise occurs immediately behind Treskavac. It occurs, relatively accurately (within applied method) speaking, between 20. April and 1. May in the spring – figures 14 and 15 or between 15. August and 27. August in the autumn – figures 16 and 17.

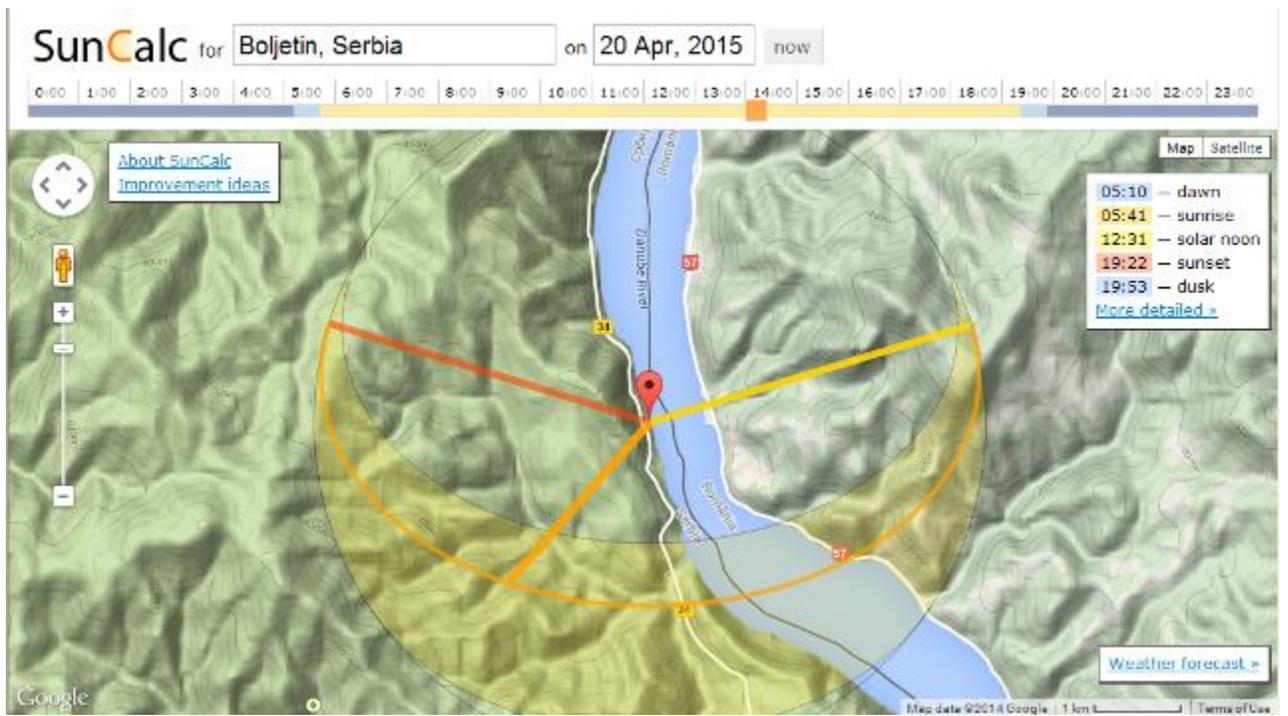

**Fig. 14**

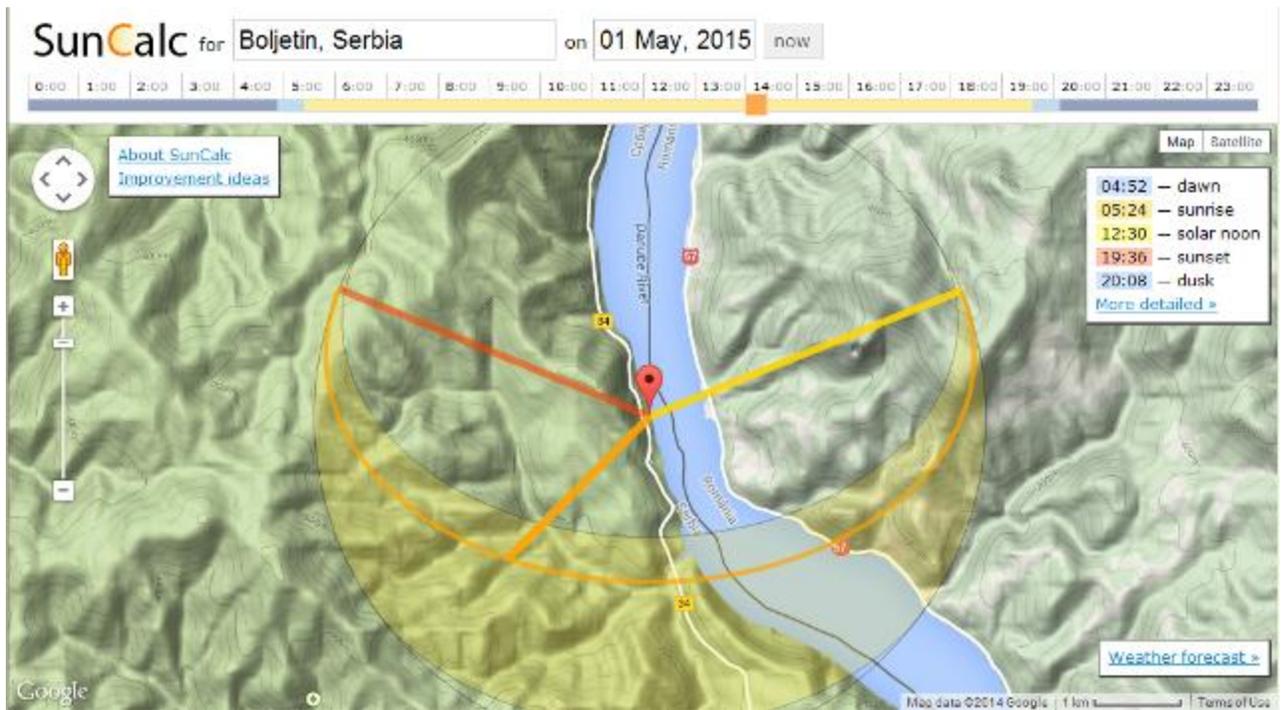

**Fig. 15**

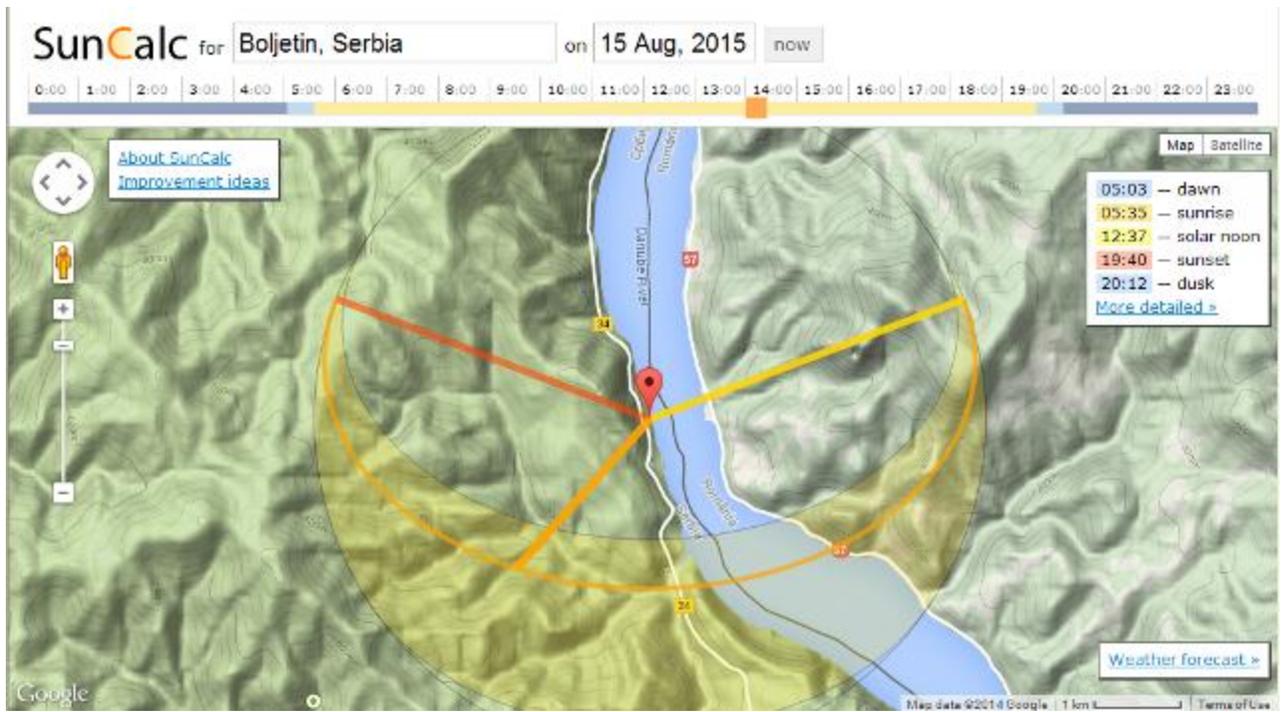

**Fig. 16**

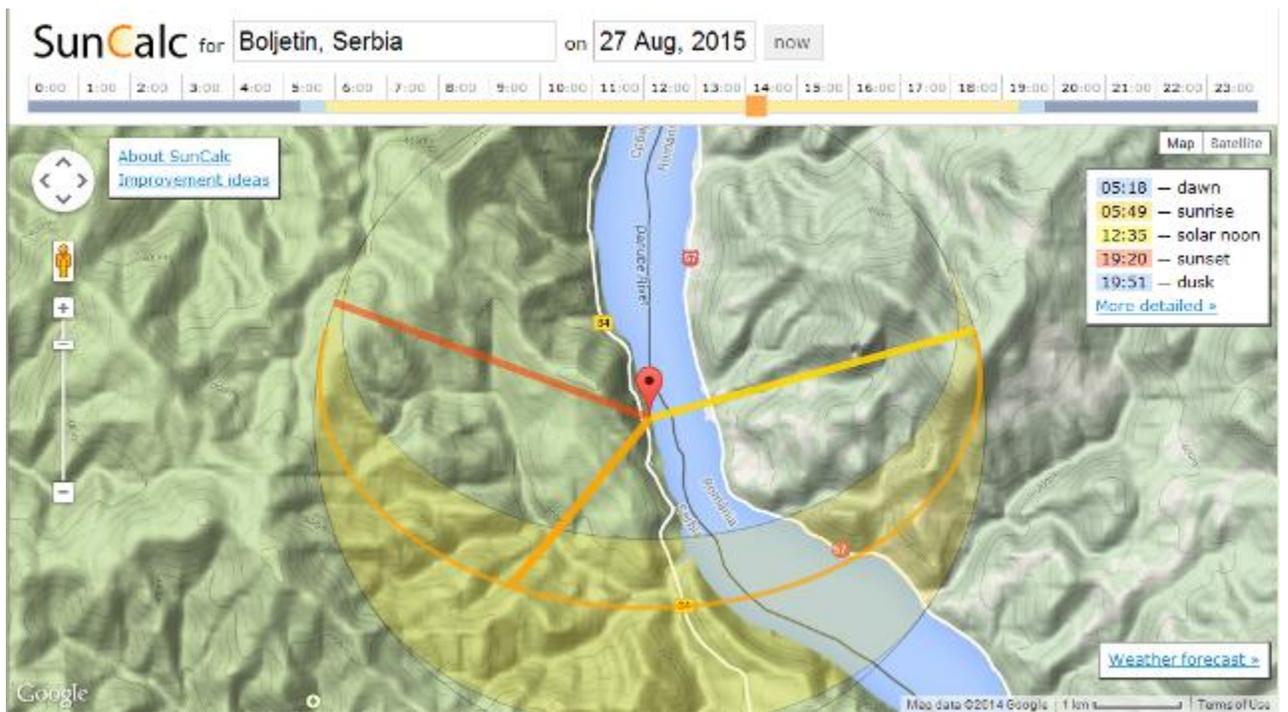

**Fig. 17**

So, with an admitable error characteristic for our method, it can be stated that sunrise beside Treskavac occurs in the Spring period between last third of the April and few first days of the May. Even if it does not represent any characteristic time interval for motion of the Sun itself, it would satisfactorily correspond to the remarkable period of the *heliacal rising* of the Pleiades constellation. Namely, during this time interval Pleiades constellation appear above the Eastern horizon very shortly before sunrise, after which, for intensive Sun shine, they become effectively non-visible. W. Kyselka states: "For the Celts of northern Europe (fifth to first centuries BC), the Pleiades' rising with the sun marked Beltane, or May Day. Since star cannot be *seen* rising with the sun, the timing of Beltane implies they had an accurate calendar." [9], 176. .
  Moreover, since it is well-known that Pleiades constellation nearly winter solstice becomes observable just after dark it is quite possible that mountain Kukuvija represented for Lepenski Vir people an implicit marker for the Pleiades constellation too.
   Finally, Pleiades constellation represents such sky object that over whole planet attracts human attention, admiration and respect from the oldest cultural epoch in the Europe, i.e. paleolithic age (Lascaux cave painting etc.) to the present [10]. For this reason, no doubt, Pleiades constellation has been excellently known to mesolithic Lepenski Vir peoples too.
  But before any final conclusion we must analyze positions of the Pleiades constellation by an Lepenski Vir observer during time interval between 20. April and 1. May. It will be done using mentioned third and fourth program, 3. http://www.spectralcalc.com/solar_calculator/solar_position.php  and 4. http://www.fourmilab.ch/cgi-bin/Yourhorizon .
  Using third program we shall now determine for Lepenski Vir observer (characterized with mentioned latitude, longitude, and URS 1 time zone) exact moment of the sunrise for some days in the time interval between 20. April and 1. May. This moment  and all other mentioned necessary characteristics, we shall further introduce in the fourth program according to which we shall obtain figure of the simulated position of the Pleiades constellation, i.e. M45 sky object by sunrise. Obtained results are presented at figures 18 – 22.

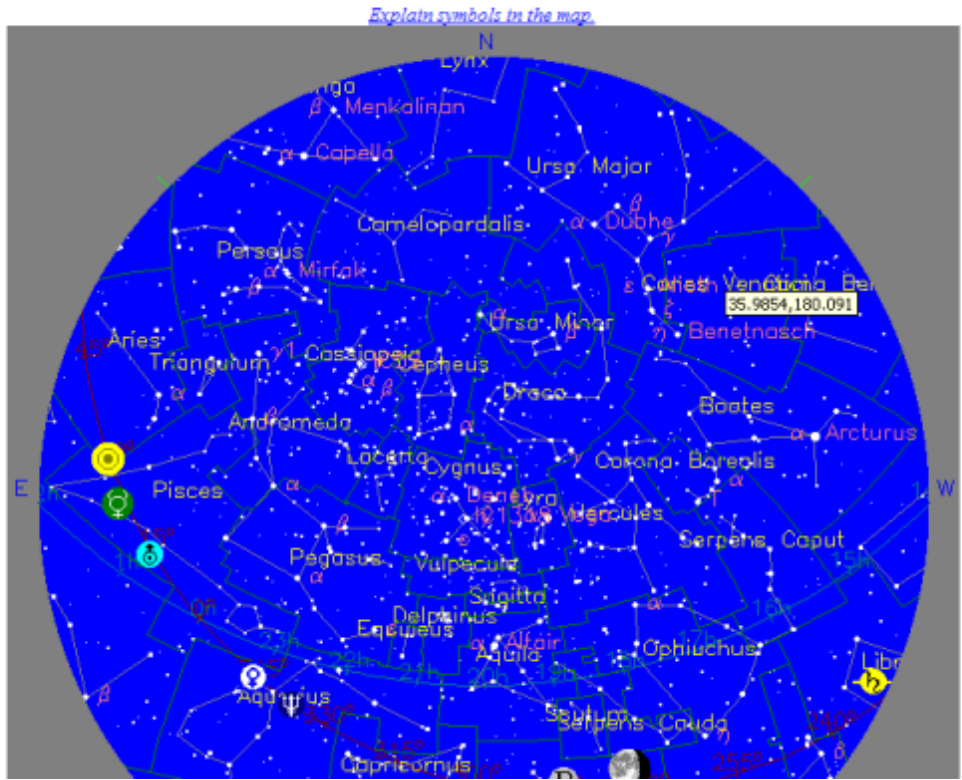

**Fig. 18**

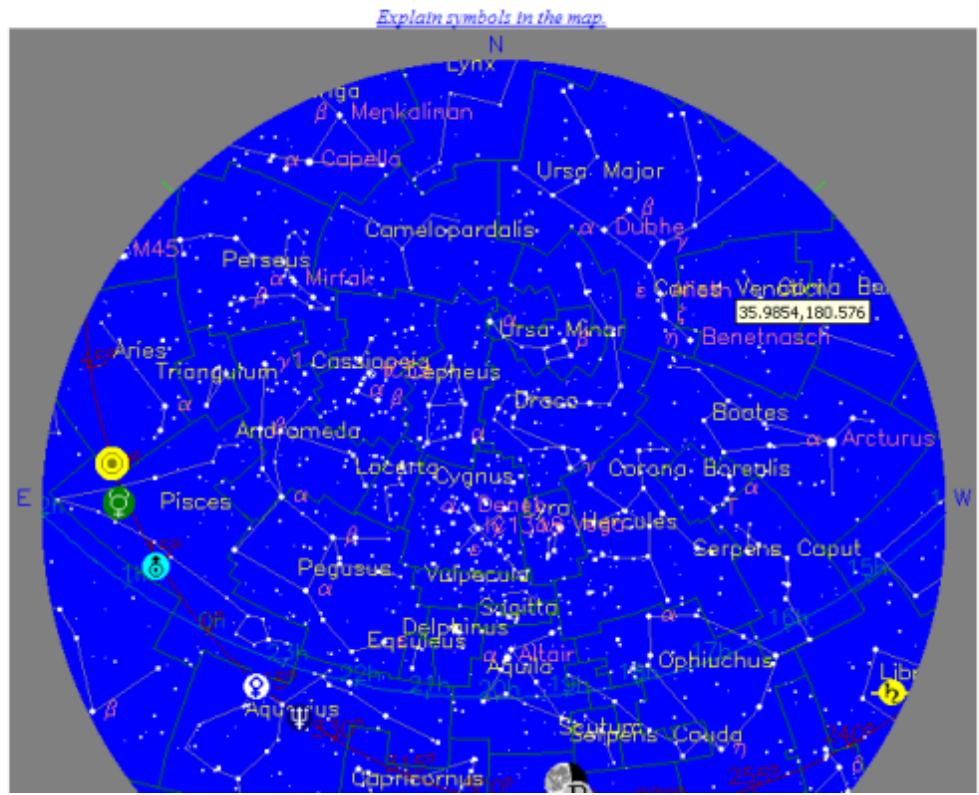

**Fig. 19**

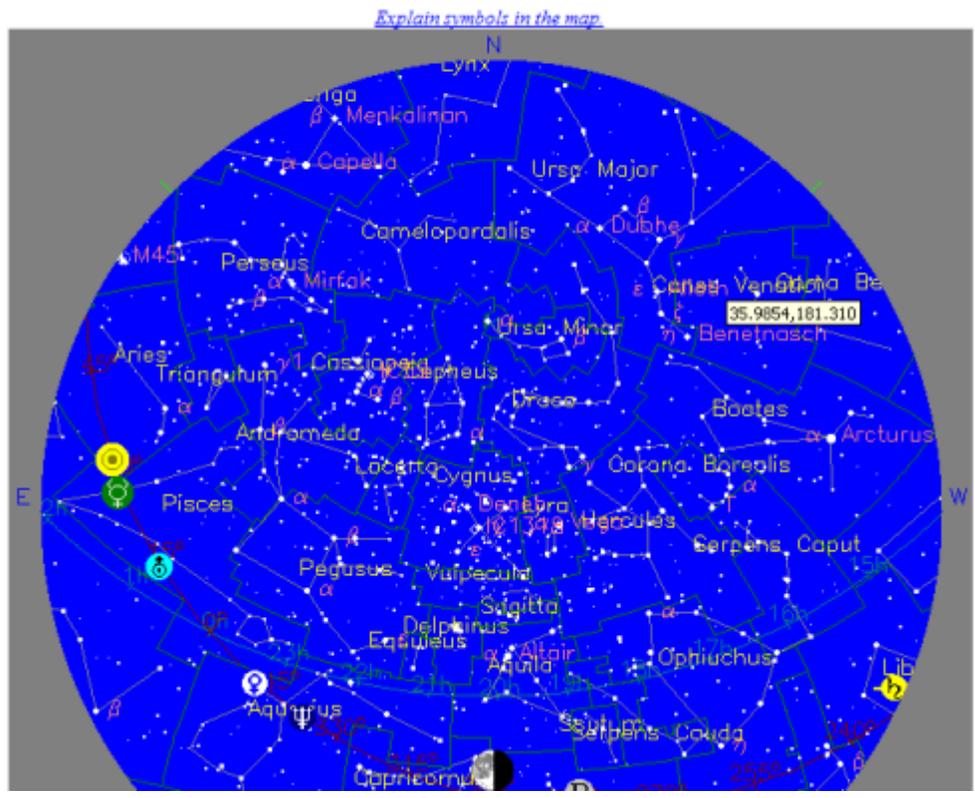

**Fig. 20**

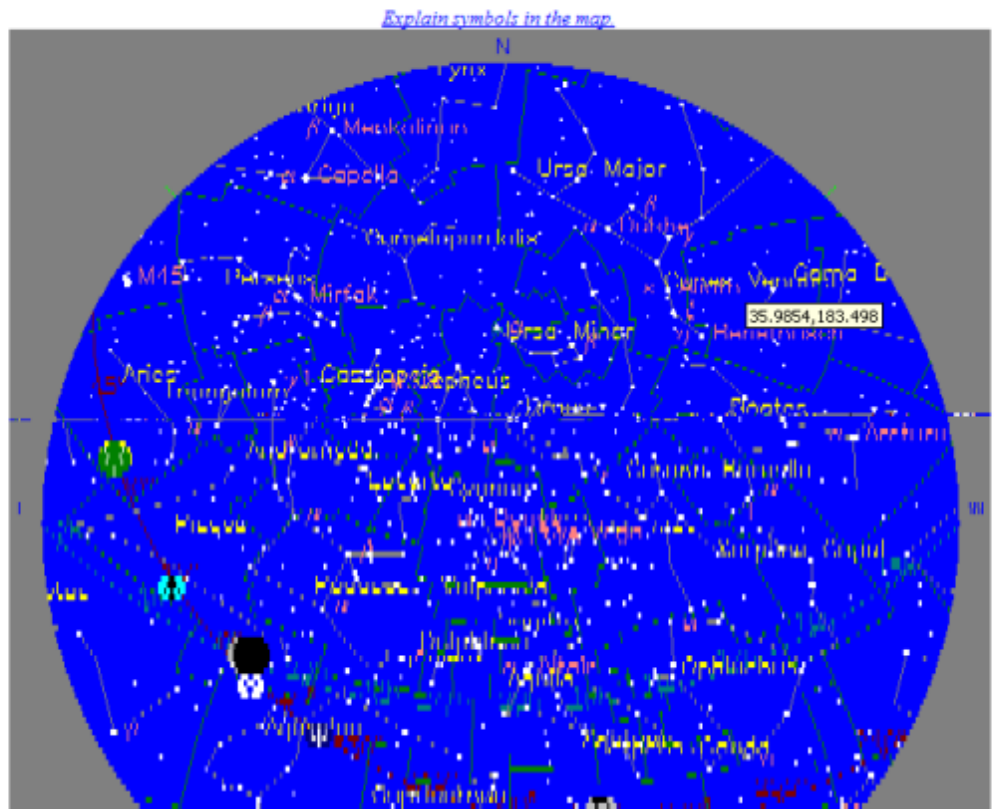

**Fig. 21**

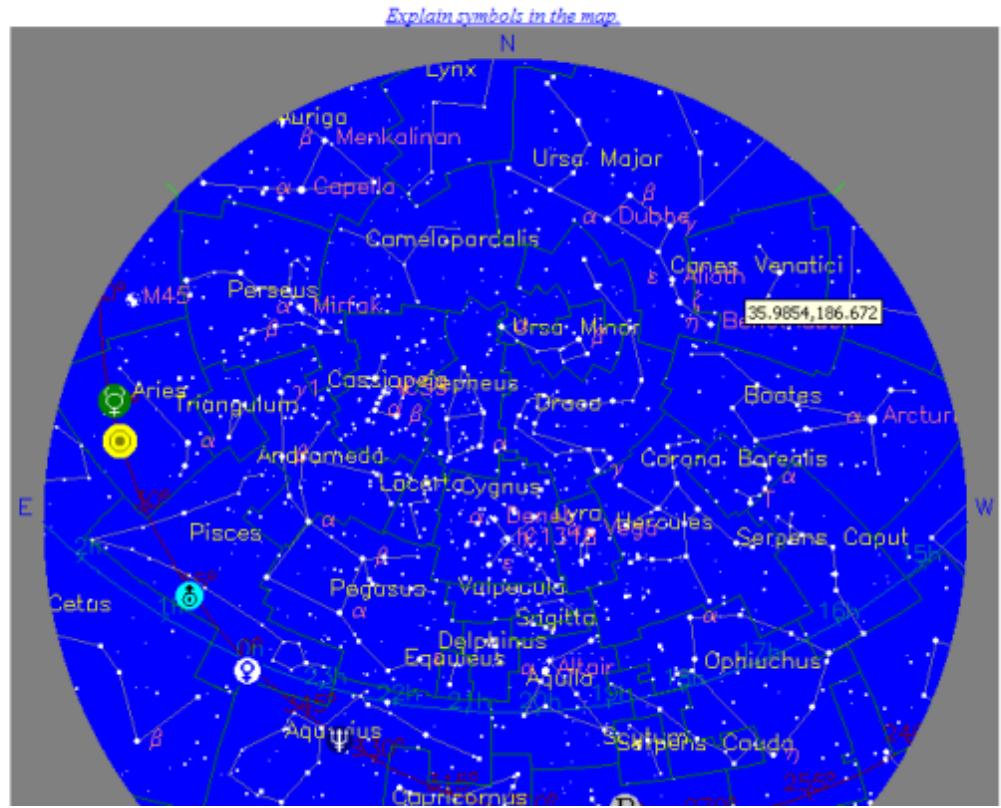

**Fig. 22**

As it is not hard to see at figure 18, Pleiades constellation appears at horizon simultaneously with sunrise at 20. April, while at later dates, figures 19 – 22, it appears somewhat later than sunrise. It practically means that Pleiades constellation becomes effectively unobservable in the same day when Sun becomes to rise exactly behind Treskavac in the Spring. (Before this day heliacal rising of the Pleiades occurs for Lepenski Vir observer.) All this represents a certainly determined event which would be used for a primitive but accurate calendar constitution.

All this opens a very probable possibility according to which Lepenski Vir was really an ancient Sun and Pleiades constellation (proto) observatory too with Treskavac as a (natural) marker.

3. Conclusion

In conclusion we can only repeat and point out the following. In this work we testify using simulation programs some old hypotheses according to which remarkable mesolithic village Lepenski Vir (6500 – 5500 BC) at the right (nearly west) Danube riverside in the Iron gate in Serbia was an ancient (one of the oldest) Sun (proto) observatory. We use method recently suggested by A. C. Sparavigna, concretely we use at internet simply and freely available software or local Sun radiation direction simulation and calculation computer programs. In this way we obtain and discuss figures of the sunrise in the Lepenski Vir during winter and summer solstice and spring and autumnal equinox in relation to position of the mountains, especially Treskavac (Trescovat in Romanian) and Kukuvija at left (nearly east) Danube riverside (in Romania). While mountain Kukuvija represents really the natural marker for the Sun in date of the winter solstice, mountain Treskavac, in despite to usual opinions, does not represent any real marker for the Sun in date of the summer solstice. Sun rises exactly behind Treskavac, relatively exactly (in the domain of used methods) speaking, between 20. April and 1. May. It more or less corresponds to year period when heliacal rising of the Pleiades constellation occurs, which by many ancient cultures, e.g. Celts of northern Europe, denotes very beginning of the year. Really, in common with some other facts obtained using at internet simply and freely available stellarium software or sky objects position simulation programs, we demonstrate that for the Lepenski Vir observer heliacal rising of the Pleiades ends exactly at 20. April and this observationally accurately determined event can be used for very beginning of the year. All this opens a very probable possibility according to which Lepenski Vir was really an ancient (one of the oldest) Sun and Pleiades constellation (proto)observatory too with Treskavac as a (natural) marker.